# Nonlinear Multi-Objective Flux Balance Analysis of the Warburg Effect

Yi Zhang* and Daniel Boley

**Abstract**—Due to its implication in cancer treatment, the Warburg Effect has received extensive in silico investigation. Flux Balance Analysis (FBA), based on constrained optimization, was successfully applied in the Warburg Effect modelling. Yet, the assumption that cell types have one invariant cellular objective severely limits the applicability of the previous FBA models. Meanwhile, we note that cell types with different objectives show different extents of the Warburg Effect. To extend the applicability of the previous model and model the disparate cellular pathway preferences in different cell types, we built a Nonlinear Multi-Objective FBA (NLMOFBA) model by including three key objective terms (ATP production rate, lactate generation rate and ATP yield) into one objective function through linear scalarization. By constructing a cellular objective map and iteratively varying the objective weights, we showed disparate cellular pathway preferences manifested by different cell types driven by their unique cellular objectives, and we gained insights about the causal relationship between cellular objectives and the Warburg Effect. In addition, we obtained other biologically consistent results by using our NLMOFBA model. For example, augmented with the constraint associated with inefficient mitochondria function or limited substrate, NLMOFBA predicts cellular pathways supported by the biology literature. Collectively, NLMOFBA can help build a complete understanding towards the Warburg Effect in different cell types.

**Keywords**: Flux balance analysis, Multi-objective optimization, Nonconvex optimization, Theoretical biology

## 1 INTRODUCTION

**Background.** Cancer cells use fermentation pathway in addition to respiration pathway for energy (ATP) production albeit fermentation pathway lowers the ATP yield (i.e. the number of ATP generated per substrate consumed). This phenomenon, first proposed by German biochemist Otto H. Warburg in 1920s, was named the Warburg Effect (Warburg, 1956). Warburg Effect is commonly observed in cancer cells and healthy proliferating cells (Liberti and Locasale, 2016; Sun et al., 2019). Generally, healthy nonproliferating cells do not show the Warburg Effect (Vander Heiden et al., 2009), but there are exceptions such as striated muscle cells and Sertoli cells (Oliveira et al., 2014; Schuster et al., 2015b).

Due to the significance of the Warburg Effect in cancer, many computational models were proposed to explain the complicated cause of the Warburg Effect (Schuster et al., 2015b; Shan et al., 2018; Shestov et al., 2014). One technique, flux balance analysis (FBA), is a constrained optimization technique generally based on linear programming (Himmel & Bomble, 2020; Orth et al., 2010). FBA assumes that cells, subjected to cellular and environmental constraints, have optimization objectives resulted from the evolutionary pressure. In FBA, the cellular objective is expressed as the objective function while the constraints are expressed as a set of equality and inequality relations. Different from the dynamic simulation using coupled ordinary differential equations, FBA is only concerned with the steady state chemical fluxes inside a biological network. As a result, FBA requires little information about the enzyme kinetics and metabolite concentrations for simulation. Prior attempts to simulate the Warburg Effect by FBA include the use of a large-metabolic network with >3000 reactions and a minimal model including three key reactions (Möller et al., 2018; Schuster et al., 2015a; Shlomi et al., 2011). Although powerful, the minimal model by Schuster et al. (2015a) is only applicable if the cells have one single objective to maximize the ATP production rate (i.e. the total number of ATPs generated from the available cellular resource and substrates).

In fact, cell types can have multiple objectives, and different cell types may have different sets of objectives (Barclay, 2017; Oh et al., 2009; Pfeiffer et al., 2001; Pfeiffer & Bonhoeffer, 2002; Vera et al., 2003). Certain cell types (e.g. healthy proliferating cells and straited muscle cells) demand a high ATP production rate for their growth or functioning (Barclay, 2017; Vander Heiden et al., 2009). From the perspective of the evolutionary game theory, cell types have the objective to maximize the ATP production rate when they are competing against other cells for the limited energy resource (Pfeiffer et al., 2001; Pfeiffer & Bonhoeffer, 2002). For example, cancer cells could have the objective to maximize the ATP production rate when invading the healthy cells. ATP yield is another common energy aspect of cellular pathways (Libretexts, 2020). ATP yield characterizes the cost-effectiveness of energy production while ATP rate characterizes the amount of energy production. Game theory suggests that some cell types (e.g. healthy nonproliferating cells) maximize the ATP yield when cooperating with each other to use the limited substrate in the most efficient manner (Pfeiffer et al., 2001; Pfeiffer &

---

- *The authors are with the Department of Computer Science & Engineering, University of Minnesota - Twin Cities, Minneapolis, MN 55455. E-mail: zhan2854@umn.edu, Boley@umn.edu.*
- *Corresponding Author*
- *Code available at https://github.com/YiZhan2854/NLMOFBA.*

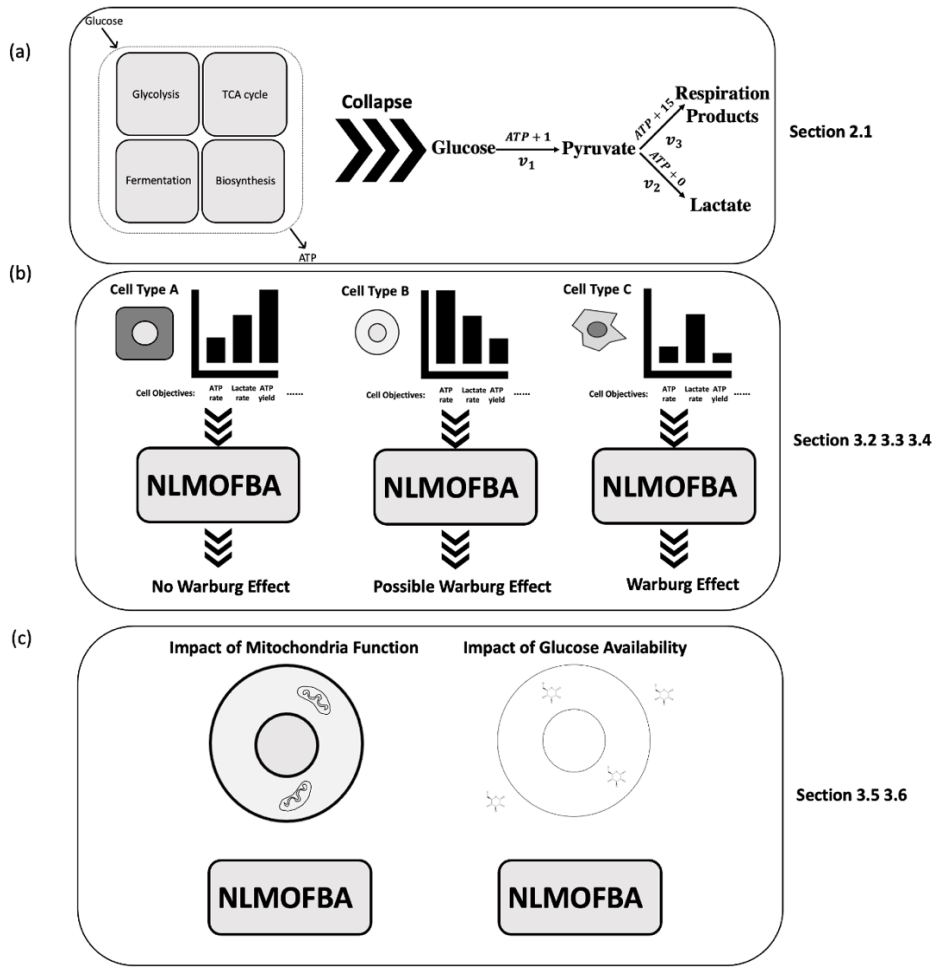

Fig. 1. A high-level explanation of the paper. (a) We collapse a large metabolic network pertinent to the Warburg Effect to a minimal one by only selecting the central pathways. The minimal metabolic network is used throughout this paper. (b) We vary the extents of three cellular objectives (ATP rate, lactate rate, ATP yield) to model different cell types. For each cell type, we take a linear combination of its cellular objectives as the objective function of the corresponding optimization problem, and we find the optimal solution by our computation model NLMOFBA to determine the Warburg Effect extent. By following these steps, we explain different extents of the Warburg Effect in different cell types. (c) Using NLMOFBA, we also model more biological conditions and their impact on the Warburg Effect. Note that each part of Fig.1 corresponds to certain section(s) in this paper, and the section numbers are shown.

Bonhoeffer, 2002). Another objective is the production of lacate, which is an important molecule involved in many cellular processes (Archetti, 2014; Gatenby 1995).

We also note that different cell types show different extents of the Warburg Effect (Sun et al., 2019; Vander Heiden et al., 2009; Warburg, 1956). Clearly, cancer cells and healthy proliferating cells are more likely to show the Warburg Effect than healthy nonproliferating cells. Thus, in this work, we seek the answer to the following theoretical biology question: can different Warburg Effect extents in diverse cell types be explained by different cellular objectives? There are some works aiming to correlate the Warburg Effect to cell objectives. For example, Archetti (2014) and Gatenby (1995) explained the Warburg Effect in cancer cells by stating that cancer cells gain growth advantages by generating lactate to "toxify" the healthy cells. Also, the demand for abundant lactate was used to explain the observed Warburg Effect in Sertoli cells (Oliveira et al., 2014).

Here, we attempt to connect the Warburg Effect to three cell objectives (i.e. ATP production rate, ATP yield, and lactate production rate) by augmenting the previous minimal model (Möller et al., 2018; Schuster et al., 2015a).

To take different cell objectives into account in our model, we resort to multi-objective optimization. Multi-objective optimization is widely applied to many domains such as economics. There are many methods to solve multi-objective optimization, and they can be generally divided to four categories: no-preference, priori, posteriori, and interactive. A common method, scalarizing, belongs to priori methods, and it combines multiple objectives into a single objective in an optimization problem. In this work, we specifically leverage linear scalarization which takes a linear combination of the weighted objectives as one objective function (Gunantara, 2018). In addition, we systematically vary the weights of different cellular objectives to investigate the cellular pathway preferences of disparate cell types. This approach is similar to varying weights in the loss function of soft margin classification (Cortes & Vapnik, 1995). Note that the ATP yield objective term is nonlinear. Thus, we denote our model as Nonlinear Multi-Objective FBA (NLMOFBA). The nonlinear ATP yield renders the entire objective function nonconvex. Compared to convex optimization, nonconvex optimization is generally more challenging to solve.

**Our contribution:** Our work makes four major contributions. First, we adopt multi-objective optimization approach to investigate the Warburg Effect, and overcome the limitation of relevant works that assume a single cell type and a single objective. By doing so, we successfully associate different Warburg Effect extents to different cellular objectives. Second, we solve the associated nonconvex optimization problem via a customized searching method with reduced computational cost. Third, we derive and prove an interface equation that captures the impact of different cellular parameters on cell behaviors. Last, our model can output many key biological results consistent with the literature, including the complicated ones.

**Outline.** The remaining paper is organized as the follows: Section 2 introduces the metabolic network, the nonlinear programming system and the implementation details of NLMOFBA; in Section 3.1, we construct a cellular objective map for selected cell types; in Section 3.2, we run NLMOFBA to unveil the cellular pathways in different cell types; in Section 3.3 and 3.4, we investigate the impact of key cellular parameters on the modelling results; in Section 3.5 and 3.6, we use NLMOFBA to reproduce key biological results of the Warburg Effect; and in Section 4, we show that NLMOFBA results are consistent with the established Warburg Effect theory. A high-level understanding of this paper is shown in Fig. 1.

## 2 METHODS

### 2.1 Minimal Metabolic Network

We first collapse a large metabolic network to a minimal one for modelling the Warburg Effect (see Fig. 1 upper right corner for the minimal metabolic network). Note that the main goal of this paper is to provide a **theoretical** explanation of the Warburg Effect instead of rigorously modeling the detailed cellular mechanisms. According to Occam's razor, in theoretical work, the simplest explanation is usually the best one to interpret complicated phenomenon. In addition, the minimal metabolic network used in this work was previously proposed by theoretical biology experts (Möller et al., 2018; Schuster et al., 2015a), and we mainly render their idea applicable to the scenarios with multiple cell types and multiple objectives by leveraging more complicated computation. Furthermore, we find at least two specific advantages of using the minimal metabolic network in this work. First, the impact of parameter selection and constraint setting on the model output is more obvious; thus, the model has high interpretability, which is particularly important in theoretical biology aiming to explain a phenomenon. Second, a smaller metabolic network leads to faster execution of the computation program with high time complexity. Thus, the use of the minimal metabolic network in our work is well justified.

In our minimal metabolic network, there are simply glycolysis reaction, fermentation reaction, and respiration reaction. The glycolysis reaction refers to the conversion from glucose to pyruvate. The fermentation reaction refers to the conversion from pyruvate to lactate. The respiration reaction refers to the complete oxidation of pyruvate through the tricarboxylic acid (TCA) cycle. Respiration pathway or fermentation pathway includes the corresponding reaction plus the glycolysis reaction.

Symbols $v_1$, $v_2$, and $v_3$ denote the reaction rates of glycolysis, fermentation, and respiration respectively. Stoichiometries of all reactants and products are set in a way such that the stoichiometry coefficient of pyruvate is always 1. This is different from the convention used by Schuster et al. (2015a), who kept the stoichiometry coefficient of glucose at 1.

### 2.2 Nonlinear Programming

Section 2.2 provides the final version of the nonlinear programming problem, followed by the derivation of the objective function, and the explanation of the constraints.

**Final nonlinear programming system:**

**Minimize:**
$$F_{objective} = -32av_1 - (b - 30a)v_2 + 210(1 - a - b)\frac{v_2}{v_1}$$
(objective function) (1)

**Subject to:**
$$\alpha_1 v_1 + \alpha_2 v_2 + \alpha_3 (v_1 - v_2) \leq \mathcal{T}$$
(total enzyme resource constraint) (2)

$$v_1 \geq 0 \quad \text{(nonnegative glycolysis rate) (3)}$$

$$v_2 \geq 0 \quad \text{(nonnegative fermentation rate) (4)}$$

$$v_1 - v_2 \geq 0 \quad \text{(nonnegative respiration rate) (5)}$$

**Objective function derivation.** Three cellular objectives are maximizing ATP production rate, maximizing lactate production rate and maximizing ATP yield (Archetti, 2014; Barclay, 2017; Gatenby, 1995; Libretexts, 2020; Oliveira et al., 2014; Pfeiffer & Bonhoeffer, 2002; Pfeiffer et al., 2001; Vander Heiden et al., 2009). Combining three objectives results in the initial form of the objective function:

**Maximize:**
$$an_1 v_{ATP} + bn_2 v_{lactate} + cn_3 Y_{ATP}. \tag{6}$$

Symbols $a$, $b$, and $c$ are the respective objective weights of $n_1 v_{ATP}$ (ATP production rate objective term), $n_2 v_{lactate}$ (lactate production rate objective term), and $n_3 Y_{ATP}$ (ATP yield objective term). Symbols $v_{ATP}$, $v_{lactate}$, and $Y_{ATP}$ are ATP production rate, lactate production rate and ATP yield respectively. They are determined by:

$$v_{ATP} = m_1 v_1 + m_3 v_3 \quad (7)$$

$$v_{lactate} = v_2 \quad (8)$$

$$Y_{ATP} = \frac{v_{ATP}}{(v_1/2)} = 2(m_1 + m_3 v_3/v_1). \quad (9)$$

Fermentation does not generate ATP. Thus, $v_{ATP}$ is the sum of the ATP generated from glycolysis and respiration (see (7)). $m_1$ and $m_3$ quantify the ATP production, and they are kept at 1 and 15 respectively because one glucose generates 2 and approximately 30 ATP via glycolysis and respiration respectively (Shlomi et al., 2011; Yetkin-Arik et al., 2019). Lactate generation rate is equal to the fermentation rate because only fermentation generates lactate (see (8)). ATP yield is the ATP produced per glucose consumed (see (9)). $n_1$, $n_2$, and $n_3$ are constant coefficients to render the maximal possible changes of three objective terms roughly similar in magnitude in the base case (Section 3.2):

$$n_1 \max(\Delta v_{ATP}) \approx n_2 \max(\Delta v_{lactate}) \approx n_3 \max(\Delta Y_{ATP}) \quad (10)$$

Constants $n_1$, $n_2$ and $n_3$ are set to 2, 1, and 7 respectively, and these values are used throughout the paper. The constant coefficients render the simulation results more meaningful.

By (7), (8), (9) and the constant values, we update the objective function in (6) with:

**Maximize:**

$$2a(v_1 + 15v_3) + bv_2 + 7c(2 + \frac{30v_3}{v_1}). \quad (11)$$

We keep the sum of $a$, $b$, and $c$ at 1. Thus, $c$ is:

$$c = 1 - a - b \quad (12)$$

Because of the steady state assumption in FBA, the fluxes of pyruvate are always balanced (i.e. $v_1 = v_2 + v_3$). Thus, $v_3$ is:

$$v_3 = v_1 - v_2 \quad (13)$$

After replacing $c$ and $v_3$ in (11) by (12) and (13), we update the objective function in (11) with:

**Maximize:**

$$32av_1 + (b - 30a)v_2 - 210(1 - a - b)\frac{v_2}{v_1} \\ +14(1 - a - b)(1 + 15). \quad (14)$$

The last term is a constant when $a$ and $b$ are specified. Thus, the last term does not impact the optimal values of $v_1$ (denoted as $v_{1,opt}$) and $v_2$ (denoted as $v_{2,opt}$). After eliminating the constant term and converting the objective function to its minimization form, we obtain the final version of the objective function in (1).

**Constraints.** Constraint (2) is due to the limited total cellular enzyme resource $\mathcal{T}$ (Müller et al., 2014). Symbols $\alpha_1$, $\alpha_2$ and $\alpha_3$ denote the enzyme cost of $v_1$, $v_2$ and $v_3$ respectively. Variable $v_3$ is replaced by $v_1 - v_2$ due to the pyruvate mass balance. The value of $\alpha_3$ should be much higher than $\alpha_1$ and $\alpha_2$ because respiration involves many steps (Möller et al., 2018; Schuster et al., 2015a). Also, using a large $\alpha_3$ is consistent with the fact that glycolysis occurs 10-100 times faster than does respiration (Liberti and Locasale, 2016).

Constraints (3), (4), and (5) are used to capture the irreversibility of three reactions. Reversible fermentation is uncommon, and thus not considered in this work. It was incorporated into a single-objective FBA model by Möller et al. (2018).

### 2.3 NLMOFBA Implementation

Nonlinear Multi-Objective Flux Balance Analysis (NLMOFBA) pseudocode is provided below.

**NLMOFBA Pseudocode:**

```
1.  PROGRAM NLMOFBA
2.  FOR a from 0 to 1 with the step size of 0.02
3.    FOR b from 0 to 1 with the step size of 0.02
4.      c = 1 - a - b
5.      IF c < 0
6.        Mark as invalid on the objective map
7.        Continue with the next iteration
8.      END IF
9.      Discretize the boundary of the feasible region
10.     Calculate the objective values at all discretization points
11.     Calculate the objective values at all corner points of the feasible region
12.     Find the minimal objective value
13.     Check if any other minimum on the boundary of the feasible region
14.     Check if any other minimum inside the feasible region
15.     Fill in the location of the minimum on the objective map
16.   END FOR
17. END FOR
18. END
```

To model different cell types, we vary $a$ and $b$ in the objective function, and $c$ is determined by (12). Our implementation could result in negative $c$. When it occurs, we mark the corresponding objective function as "invalid" and skip the iteration. We call the figures with the objective weights as their axes "objective maps" (e.g. Fig. 2 and 4).

For each valid objective function, the optimal point ($v_{1,opt}$, $v_{2,opt}$) is determined by solving the corresponding nonlinear programming problem. The objective function in (1) is nonconvex over the feasible region unless $c = 0$ (proof in Appendix Section II-1). To solve the nonconvex optimization problem, we search for the optimum over the feasible region exhaustively through discretization. Although robust, exhaustive search is computationally expensive, especially when the feasible region has a high dimension. The feasible region in our problem is two-dimensional. Exhaustively searching the two-dimensional feasible region results in a time complexity of $O(n^2)$, where $n$ is the number of discretization points used for each dimension. Considering the two outer loops required by the multi-objective approach, the exhaustive search will make the program computationally expensive.

Through the mathematical analysis of this specific problem, we obtain the following result:

TABLE 1
SYMBOL INTERPRETATIONS

| Symbols | Interpretations |
|---|---|
| $a$ | Weight of ATP production rate in the objective function |
| $b$ | Weight of lactate generation in the objective function |
| $c$ | Weight of yield in the objective function |
| $v_1$ | Glycolysis reaction rate |
| $v_2$ | Fermentation reaction rate |
| $v_3$ | Respiration reaction rate |
| $m_1$ | Glycolysis ATP output |
| $m_3$ | Respiration ATP output |
| $\alpha_1$ | Enzyme cost for glycolysis reaction |
| $\alpha_2$ | Enzyme cost for fermentation reaction |
| $\alpha_3$ | Enzyme cost for respiration reaction |
| $\mathcal{T}$ | Total cellular enzyme resource |
| $X_3$ | Upper bound on respiration rate |
| $V_{glucose}$ | Glucose availability |
| $v_{ATP}$ | Total ATP production rate |
| $Y_{ATP}$ | ATP yield per glucose |
| $v_{lactate}$ | Total lactate production rate |
| $f_{fermentation}$ | Fraction of fermentation |
| $F_{objective}$ | Objective function |
| $\Delta$ | Difference |

TABLE 2
PARAMETER SETTINGS IN FIGURES AND SECTIONS

| Symbols | Fig. 4 (Sec 3.2) | Fig. S1a (Sec 3.2) | Fig. S1b (Sec 3.2) | Fig. S1c (Sec 3.2) | Fig. S1d (Sec 3.2) |
|---|---|---|---|---|---|
| $a$ | 0:0.02:1 | 0.5 | 0.33 | 0.15 | 0 |
| $b$ | 0:0.02:1 | 0.5 | 0.33 | 0.15 | 0 |
| $c$ | $1-a-b$ | 0 | 0.34 | 0.7 | 1 |
| $\alpha_1$ | 0.5 | 0.5 | 0.5 | 0.5 | 0.5 |
| $\alpha_2$ | 0.5 | 0.5 | 0.5 | 0.5 | 0.5 |
| $\alpha_3$ | 10 | 10 | 10 | 10 | 10 |
| $\mathcal{T}$ | 200 | 200 | 200 | 200 | 200 |
| $m_1$ | 1 | 1 | 1 | 1 | 1 |
| $m_3$ | 15 | 15 | 15 | 15 | 15 |
| $X_3$ | N/A | N/A | N/A | N/A | N/A |
| $V_{glucose}$ | N/A | N/A | N/A | N/A | N/A |
| Symbols | Fig. 5b (Sec 3.3) | Fig. 6a (Sec 3.4) | Fig. 6b (Sec 3.4) | Fig. 6c (Sec 3.4) | Fig. 6d (Sec 3.4) |
| $a$ | 0:0.02:1 | 0:0.02:1 | 0:0.02:1 | 0:0.02:1 | 0:0.02:1 |
| $b$ | 0:0.02:1 | 0:0.02:1 | 0:0.02:1 | 0:0.02:1 | 0:0.02:1 |
| $c$ | $1-a-b$ | $1-a-b$ | $1-a-b$ | $1-a-b$ | $1-a-b$ |
| $\alpha_1$ | 0.5 | Varying | 0.5 | 0.5 | 0.5 |
| $\alpha_2$ | 0.5 | 0.5 | Varying | 0.5 | 0.5 |
| $\alpha_3$ | 25 | 10 | 10 | Varying | 10 |
| $\mathcal{T}$ | 200 | 200 | 200 | 200 | Varying |
| $m_1$ | 1 | 1 | 1 | 1 | 1 |
| $m_3$ | 15 | 15 | 15 | 15 | 15 |
| $X_3$ | N/A | N/A | N/A | N/A | N/A |
| $V_{glucose}$ | N/A | N/A | N/A | N/A | N/A |
| Symbols | Fig. 6e (Sec 3.4) | Fig. 7b (Sec 3.5) | Fig. 8c (Sec 3.6) | | |
| $a$ | 0:0.02:1 | 0:0.02:1 | 0.5:0.01:1 | | |
| $b$ | 0:0.02:1 | 0:0.02:1 | 0 | | |
| $c$ | $1-a-b$ | $1-a-b$ | $1-a$ | | |
| $\alpha_1$ | 0.5 | 0.5 | 0.5 | | |
| $\alpha_2$ | 0.5 | 0.5 | 0.5 | | |
| $\alpha_3$ | 25 | 10 | 25 | | |
| $\mathcal{T}$ | Varying | 200 | 200 | | |
| $m_1$ | 1 | 1 | 1 | | |
| $m_3$ | 15 | 15 | 15 | | |
| $X_3$ | N/A | 10 | N/A | | |
| $V_{glucose}$ | N/A | N/A | 1:5:201 | | |

*The notation A:B:C denotes an array from A to C with increments of B.*

**Result 1.** The optimal point $(v_{1,opt}, v_{2,opt})$ is always at the boundary of the feasible region unless $\frac{dF_{objective}}{dv_2} = 0$ at $(v_{1,opt}, v_{2,opt})$ (See Appendix Section II-2 for the proof).

Thus, it is sufficient to search the boundary of the feasible region for $(v_{1,opt}, v_{2,opt})$ when $\frac{dF_{objective}}{dv_2} \neq 0$. Exhaustively searching the one-dimensional feasible region boundary reduces the time complexity to $O(n)$, where $n$ is the number of discretization points used for one dimension. Note that we can further reduce the computational cost based on **Result 2**:

**Result 2.** The optimal point $(v_{1,opt}, v_{2,opt})$ is always on the right side of the feasible region (example of feasible region in Fig. 3) unless $a = 0$ and $v_2 = 0$ (See Appendix Section II-3 for the proof).

We do not use **Result 2** because it does not impact the time complexity of the program. However, **Result 2** will be useful when we derive the interface equation in Section 3.4.

Thus, we discretize the feasible region boundary, and we search for the minimal $F_{objective}$ over the discretized points and the corner points of the feasible region. Next, we check the uniqueness of the optimal point $(v_{1,opt}, v_{2,opt})$ along the feasible region boundary. If there is only one optimal point along the feasible region boundary and $\frac{dF_{objective}}{dv_2}$ is nonzero at $(v_{1,opt}, v_{2,opt})$, we know immediately that the optimal point is unique over the entire feasible region by **Result 1**. We also fill the objective map with the locations of $(v_{1,opt}, v_{2,opt})$ in the feasible region for all valid objective weight combinations.

Some search methods such as golden-section search are generally more accurate (Press, William H., 2007). But in this specific study, the discretization-based search method is sufficient to result in the exact values of the optimal solutions.

The program is implemented in MATLAB. Fraction of fermentation $f_{fermentation}$ is used to quantify the Warburg Effect. $f_{fermentation}$ is defined as:

$$f_{fermentation} = \frac{v_{2,opt}}{v_{1,opt}} \qquad (15)$$

Note that $v_{1,opt}$ is the sum of $v_{2,opt}$ and $v_{3,opt}$. Thus, $f_{fermentation} = 1$ indicates that cells use 100% fermentation. $f_{fermentation} = 0$ indicates that cells use 100% respiration. Any in-between values indicate that cells use mixed respiro-fermentation. ATP production rate, lactate production rate and ATP yield at the optimal solution can be determined by plugging $v_{1,opt}$, $v_{2,opt}$ and $v_{3,opt}$ into (7), (8), and (9).

### 2.4 Simulate More Biological Conditions

We separately simulate two additional biological conditions.

Although the proposal that the Warburg Effect is completely caused by the malfunction of mitochondria has been disproved, less efficient mitochondria function is still one of the leading contribution factors that lead to the Warburg Effect (Harris & Johnson, 2019). The most obvious effect of the inefficient mitochondria function can be simulated by placing an upper bound on the rate of respiration, which occurs in the mitochondria matrix of eukaryotic cells. Mathematically, it is

$$v_3 \leq X_3 \qquad (16)$$

where $X_3$ is the upper bound of $v_3$. The variable $v_3$ is replaced by $v_1 - v_2$ during the implementation due the pyruvate mass balance.

Low glucose availability occurs in many cellular environments [4]. Thus, we investigate the impact of the glucose availability on the Warburg Effect. Limited glucose availability places an upper bound $V_{glucose}$ on $v_1$:

$$v_1 \leq V_{glucose} \qquad (17)$$

### 2.5 Parameter Value Selection

In most cases, we use the parameter values from the previous papers (i.e. Möller et al., 2018; Schuster et al., 2015a, 2015b), all of which provide justifications about the selection of these values. To make this paper self-contained, we also include justifications when these parameters appear. Interpretations of all symbols are in Table 1. Parameter settings of figures are described in their associated text, and also summarized in Table 2.

## 3 RESULTS
### 3.1 Cellular Objective Map

The fundamental FBA assumption is that cells have optimization objectives because of evolution (Orth et al., 2010). NLMOFBA is motivated by the fact that different cells could have dissimilar objectives due to their varying cellular structures and biological purposes in living entities. Here, we propose an approximated cellular objective map for four selected cell types (cancer cells, healthy proliferating cells, healthy nonproliferating cells and Sertoli cells) (Fig. 2) to facilitate the remaining discussion.

Cancer cells have heavy objectives of lactate and ATP production rates (Archetti, 2014; Gatenby, 1995; Pfeiffer & Bonhoeffer, 2002; Pfeiffer et al., 2001). Thus, their objective weight combinations are likely to stay in the region where $a + b > 0.8$ (or equivalently $c < 0.2$). Little evidence suggests that general healthy proliferating cells and healthy nonproliferating cells are "interested" in producing lactate. Thus, their objective weight combinations should stay within the region where $b < 0.1$. Healthy proliferating cells have a heavy objective of ATP production rate (Stouthamer, 1973; Vander Heiden et al., 2009). Thus, their objective weight combinations are likely to stay in the region where $a > 0.5$. Healthy nonproliferating cells have a heavy objective of ATP yield (Pfeiffer & Bonhoeffer, 2002; Pfeiffer et al., 2001). Thus, their objective weight combinations are likely to stay in the region where $c > 0.5$. Sertoli cell is an atypical healthy nonproliferating cell. It has a heavy objective of lactate production (Oliveira et al., 2014). Thus, its objective weight combinations are likely to stay in the region where $b > 0.5$.

Note that the objective weights $a$, $b$ and $c$ are used to represent the relative importance of different objectives. Their absolute values are meaningless. Also, the objective map could have more dimensions if extra objectives are incorporated. In addition, more cell types can find their locations on the map given information about their objectives.

### 3.2 Base Case

The nonlinear programming problem is listed at the beginning of Section 2.2. The parameter setting of the base case is: $\alpha_1 = \alpha_2 = 0.5$, $\alpha_3 = 10$ and $\mathcal{T} = 200$. The feasible region is shown in Fig. 3. The level curves corresponding to four objective weight combinations ($a = b = 0.5, c = 0$; $a = b = 0.33, c = 0.34$; $a = b = 0.15, c = 0.7$; $a = b = 0, c = 1$) are in Appendix (Fig. S1).

NLMOFBA in Section 2.3 is implemented. $f_{fermentation}$ values and the locations of the optimal points in the feasible region are determined for all valid objective weight combinations (Fig. 4).

With the base case parameter setting, the heavy lactate production objective (i.e. $b > 0.5$ or equivalently $b > a + c$) drives cells to use 100% fermentation, and cells with light

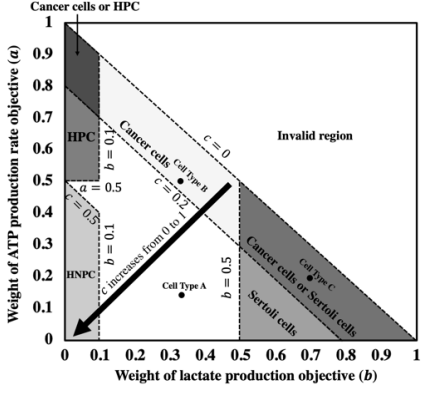

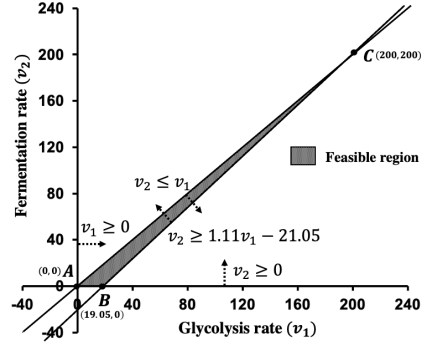

Fig. 2. Cellular objective map for selected cell types. Healthy proliferating cells and healthy nonproliferating cells are abbreviated by HPC and HNPC respectively. The objective weight $c$ of each point can be determined by $1 - a - b$. The level curves of $c$ have a slope of $-1$. The locations of three cell type examples in Fig. 1 are shown in Fig. 2.

Fig. 3. Feasible region for the constrained optimization with the base case parameter setting. Dashed arrows denote the constraint directions.

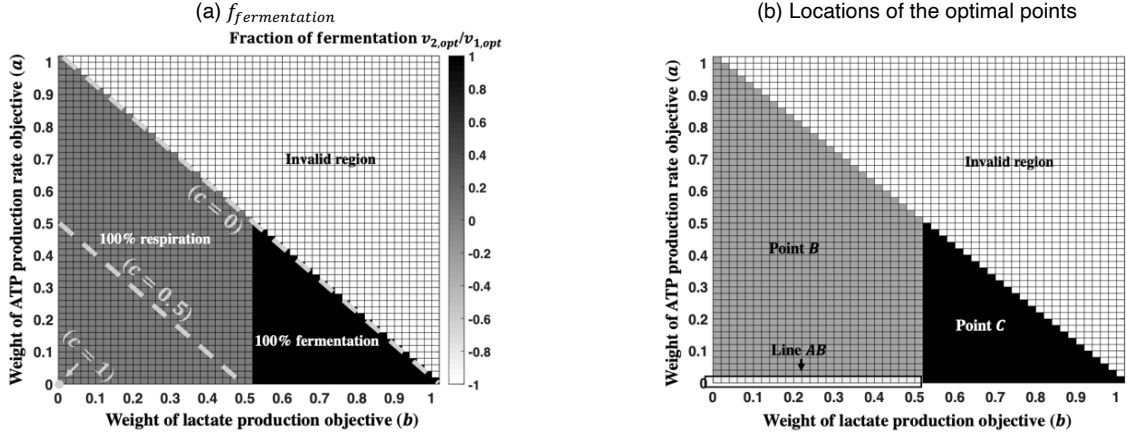

Fig. 4. (a) $f_{fermentation}$ for all valid objective weight combinations. Grey grids ($f_{fermentation} = 0$) and black grids ($f_{fermentation} = 1$) indicate 100% respiration and 100% fermentation respectively. White grids indicate the invalid region where $c < 0$. Selected levels of $c$ ($c = 0$, 0.5, or 1) are indicated by the grey dashed lines/dot. (b) Locations of the optimal points in the feasible region (Fig. 3) for all valid objective weight combinations. On the following cellular objective maps, $f_{fermentation}$ and locations of the optimal points will be included in one figure.

lactate production objective ($b < 0.5$ or equivalently $a + c > b$) use 100% respiration.

Cells with a heavy lactate production objective always prefer the fermentation pathway because only fermentation reaction generates lactate. By contrast, cells with a heavy ATP yield objective always prefer the respiration pathway because respiration maximizes the ATP yield. The effect of ATP production rate objective on the cellular operating modes depends on the parameter setting. Möller et al. (2018) showed that the respiration pathway is preferred to maximize the ATP production rate when the following condition holds:

$$\frac{m_1}{m_1+m_3} < \frac{\alpha_1+\alpha_2}{\alpha_1+\alpha_3}. \quad (18)$$

If (18) holds, respiration pathway maximizes ATP rate and ATP yield simultaneously. In other words, the objectives of ATP rate and ATP yield are nonconflicting. If (18) becomes the other way, fermentation becomes the preferred pathway to maximize the ATP production rate. In this case, ATP production rate and ATP yield become conflicting. If (18) becomes an equality relation, Constraint (2) becomes parallel to the level curves of the ATP production rate (see (7)). 100% respiration, 100% fermentation and any mixed respiro-fermentation points bound by (2) result in the maximal ATP production rate. One way for this to happen is to change $\alpha_3$ from 10 to 15.5 while keeping the other parameters unchanged. The resulting feasible region and the level curves of the ATP production rate corresponding to this parameter setting are in Appendix (Fig. S2). Thus, given limited cellular enzyme resource, the preferred pathway to maximize the ATP production rate is highly dependent on the cellular enzyme costs and numbers of generated ATP of reactions.

Inequality (18) holds in the base case. Thus, both objectives of ATP production rate and ATP yield drive cells to use the respiration pathway. The results corresponding to the conflicting objectives of ATP production rate and ATP yield will be discussed in Section 3.3.

In the base case, 100% use of fermentation always occurs at point C ($v_{1,opt} = 200$, $v_{2,opt} = 200$, $v_{3,opt} = 0$) of the feasible region (Fig. 3). Most optimal solutions leading to 100% respiration are at point B ($v_{1,opt} = 19.05$, $v_{2,opt} = 0$, $v_{3,opt} = 19.05$) (Fig. 3). Point B and point C are bounded by the total cellular enzyme resource constraint. When $a = 0$ and $b \leq 0.5$, all points along the line AB (Fig. 3) are optimal and they are not bounded by the total cellular resource constraint. For these objective weight combinations, ATP yield outweighs the lactate production rate in the objective function. Any point along the line AB will maximize the ATP yield, and there is no incentive to maximize the ATP production rate because $a = 0$.

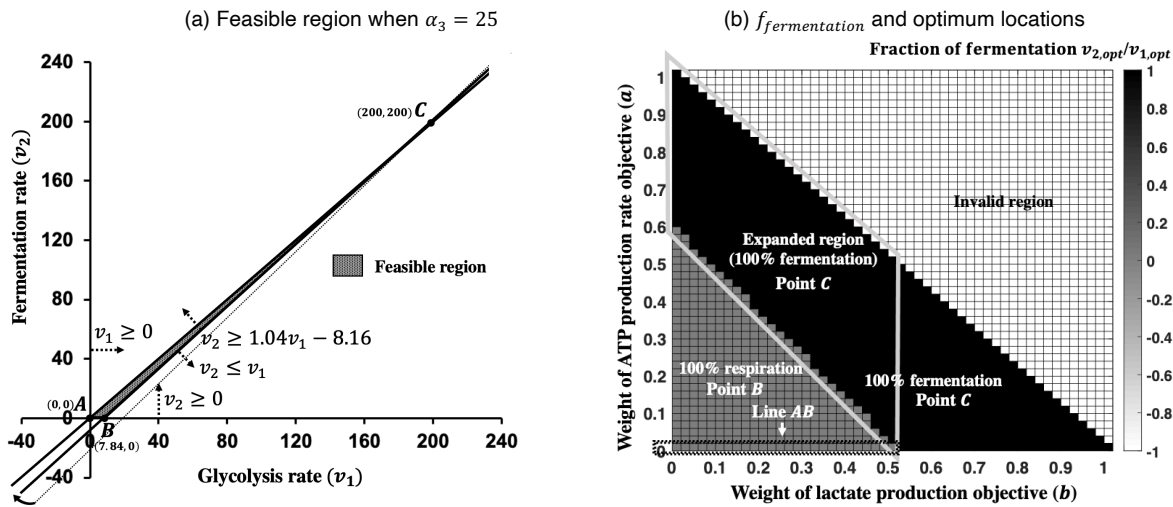

Fig. 5. (a) Feasible region for the base case variant where the objectives of ATP yield and ATP production are conflicting. Dashed line represents Constraint (2) when $\alpha_3 = 10$. It rotates to become BC as $\alpha_3$ increases from 10 to 25. (b) $f_{fermentation}$ obtained for all valid objective weight combinations when objectives of ATP yield and ATP production rate are conflicting. The black grids enclosed by the grey quadrilateral represent the expanded fermentation region due to the change in $\alpha_3$. Locations of the optimal points in the feasible region are also shown.

### 3.3 Base Case Variant

In this section, we investigate a base case variant where objectives of ATP production rate and ATP yield are conflicting (i.e. Inequality (18) changes the direction). One way for this to happen is to increase $\alpha_3$ from 10 to 25 while maintaining values of other parameters in the base case. $\alpha_3$ is an intrinsic cellular parameter, and different cell types could have different $\alpha_3$ values. Besides the objective, different values of the intrinsic cellular parameters can also be used to distinguish among cell types. The parameter setting of the base case variant is $\alpha_1 = \alpha_2 = 0.5$, $\alpha_3 = 25$ and $\mathcal{T} = 200$. Changing $\alpha_3$ alters the feasible region (Fig. 5a). The $f_{fermentation}$ and optimum locations are determined for all valid objective weight combinations (Fig. 5b).

When the objectives of ATP yield and ATP production rate are conflicting, the fermentation region in Fig. 5b expands significantly relative to the region in Fig. 4a because the fermentation pathway becomes the preferable pathway to maximize the ATP production rate. As always, cells with a heavy lactate production objective prefer to use the fermentation pathway. Thus, only cells with a heavy ATP yield objective use the respiration pathway.

### 3.4 Vary the Key Parameters

Besides $\alpha_3$, other intrinsic cellular parameters $\alpha_1$, $\alpha_2$, and $\mathcal{T}$ could also vary in different cell types. Varying cellular parameters results in the change of the interface between the fermentation and respiration regions on the cellular objective map (e.g. the interfaces between the grey and black regions in Fig. 4a and Fig. 5b). As we have already seen, changing $\alpha_3$ from 10 to 25 alters the interface (Fig. 4a and Fig. 5b). We derive the general interface equation (See Appendix Section II-4 for the derivation).

$$a = \left(\frac{\alpha_1 \mathcal{T} + \alpha_3 \mathcal{T} + 210(\alpha_1 + \alpha_3)(\alpha_1 + \alpha_2)}{30\alpha_1 \mathcal{T} + 32\alpha_2 \mathcal{T} - 2\alpha_3 \mathcal{T} - 210(\alpha_1 + \alpha_3)(\alpha_1 + \alpha_2)}\right) b - \frac{210(\alpha_1 + \alpha_3)(\alpha_1 + \alpha_2)}{30\alpha_1 \mathcal{T} + 32\alpha_2 \mathcal{T} - 2\alpha_3 \mathcal{T} - 210(\alpha_1 + \alpha_3)(\alpha_1 + \alpha_2)} \quad (19)$$

Using the interface equation, we investigate the impact of varying $\alpha_1$, $\alpha_2$, $\alpha_3$, and $\mathcal{T}$ on the interfaces of fermentation and respiration regions (Fig. 6). We confirm the validity of the interface equation by running NLMOFBA for all the parameter settings used in Fig. 6.

Increasing $\alpha_3$, decreasing $\alpha_1$, or decreasing $\alpha_2$ renders the fermentation pathway more preferable to maximize the ATP production rate (Fig. 6a, 6b and 6c). Varying $\alpha_1$, $\alpha_2$ or $\alpha_3$ has a similar effect on the expansion trend of the fermentation region. Increasing $\mathcal{T}$ always results in the expansion of the fermentation region (Fig. 6d and 6e). Varying $\mathcal{T}$ has different effects on the expansion trend of the fermentation region when ATP objectives are conflicting and non-conflicting.

### 3.5 Inefficient Mitochondria Function

As reasoned in Section 2.4, inefficient mitochondria function is modelled by placing an upper bound $X_3$ on $v_3$. The corresponding nonlinear programming problem is composed of (1), Constraints (2), (3), (4), (5), and (16). We set $X_3$ at 10. Base case values of other parameters in Section 3.2 are used ($\alpha_1 = \alpha_2 = 0.5$, $\alpha_3 = 10$, and $\mathcal{T} = 200$). The feasible region, $f_{fermentation}$ values, and locations of the optimal points in the feasible region are in Fig. 7.

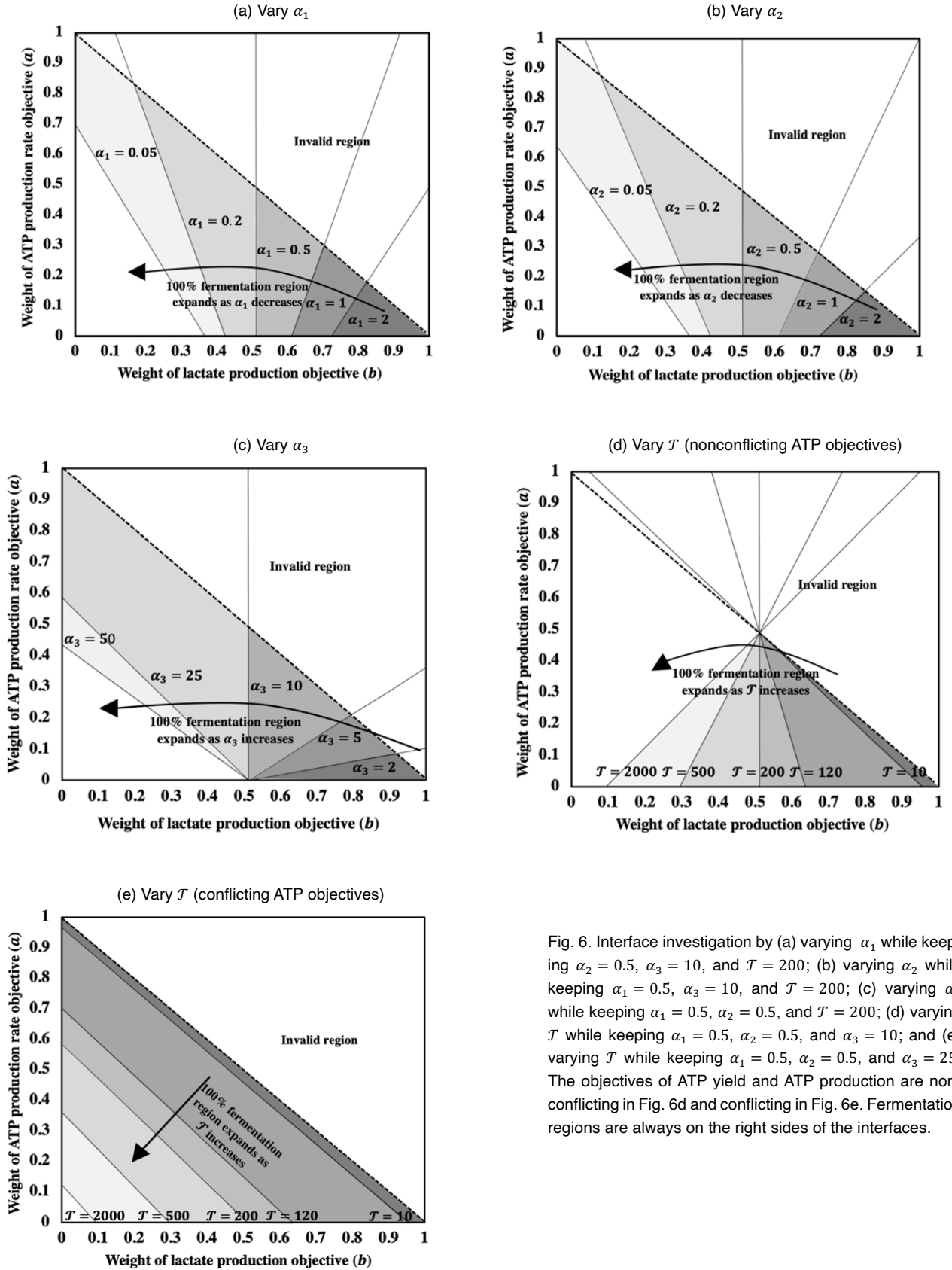

Fig. 6. Interface investigation by (a) varying $\alpha_1$ while keeping $\alpha_2 = 0.5$, $\alpha_3 = 10$, and $\mathcal{T} = 200$; (b) varying $\alpha_2$ while keeping $\alpha_1 = 0.5$, $\alpha_3 = 10$, and $\mathcal{T} = 200$; (c) varying $\alpha_3$ while keeping $\alpha_1 = 0.5$, $\alpha_2 = 0.5$, and $\mathcal{T} = 200$; (d) varying $\mathcal{T}$ while keeping $\alpha_1 = 0.5$, $\alpha_2 = 0.5$, and $\alpha_3 = 10$; and (e) varying $\mathcal{T}$ while keeping $\alpha_1 = 0.5$, $\alpha_2 = 0.5$, and $\alpha_3 = 25$. The objectives of ATP yield and ATP production are nonconflicting in Fig. 6d and conflicting in Fig. 6e. Fermentation regions are always on the right sides of the interfaces.

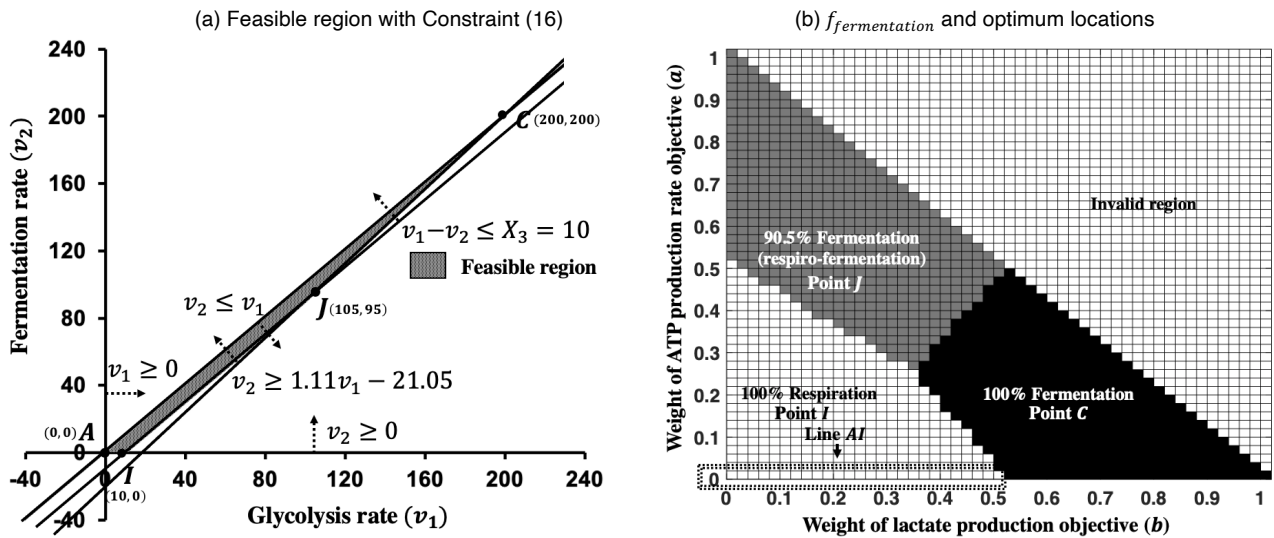

Fig. 7. (a) Feasible region after placing an upper bound on $v_3$. (b) $f_{fermentation}$ and locations of the optimal points in the feasible region.

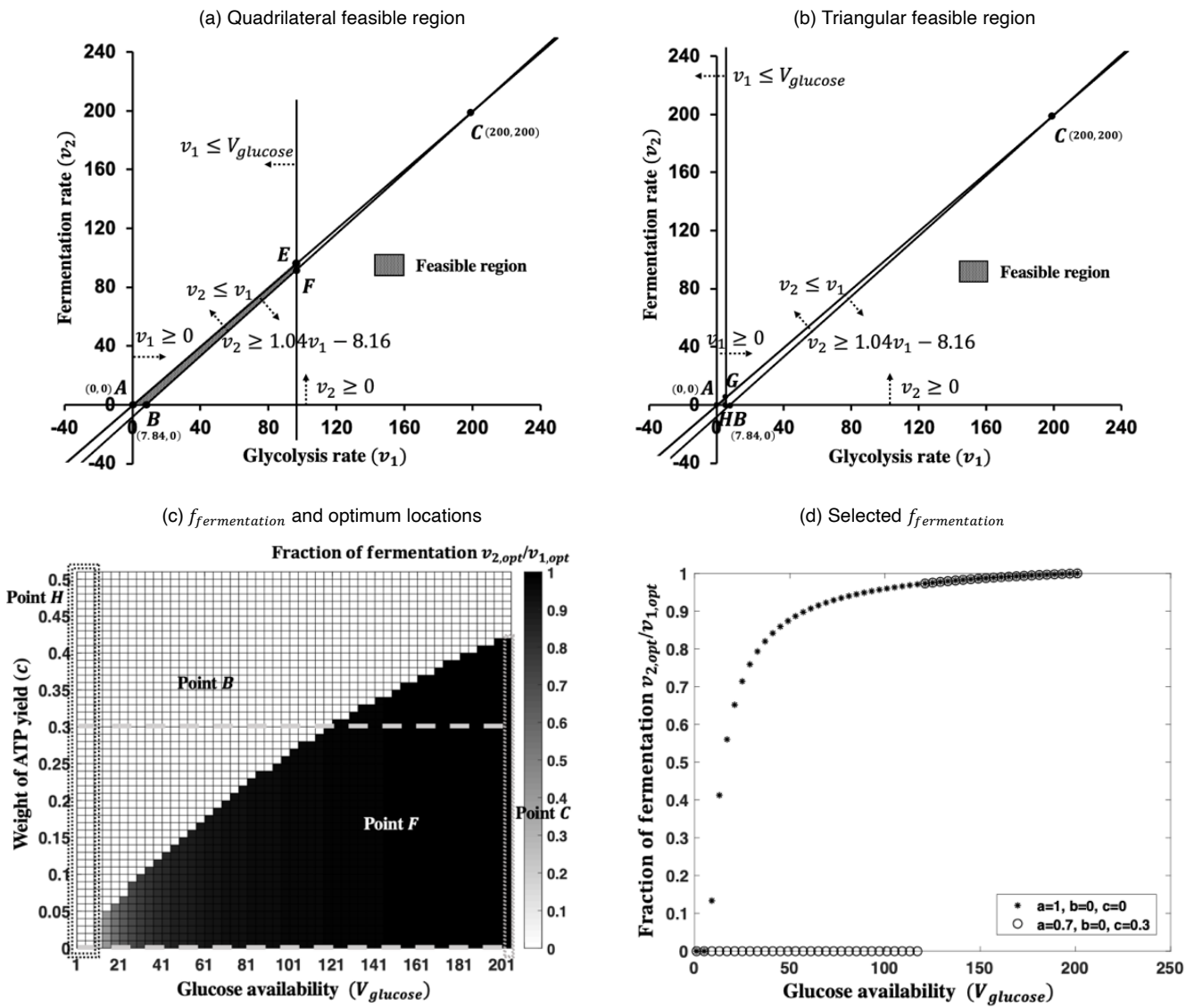

Fig. 8. (a) and (b) Possible feasible regions. (c) $f_{fermentation}$ and locations of the optimal points in the feasible region for different healthy cells subjected to different glucose availabilities. d) $f_{fermentation}$ for selected objective weight sets.

Compared to Fig. 4a, imposing an upper bound on $v_3$ results in the expansion of 100% fermentation region and the formation of a large mixed respiro-fermentation region in Fig. 7b. Mixed respiro-fermentation occurs when the optimal solution is point J in Fig. 7a.

### 3.6 Limited Glucose Availability

Glucose, the most important energy substrate, could be limited in some cellular environments (Schuster et al., 2015a; Vander Heiden et al., 2009). To simulate this effect, we place an upper bound $V_{glucose}$ on $v_1$ (Constraint (17)). The corresponding nonlinear programming problem is composed of (1), Constraints (2), (3), (4), (5), and (17). We investigate the behavior of healthy proliferating cells and healthy nonproliferating cells when subjected to different glucose availabilities by varying $V_{glucose}$ from 1 to 201. Generally, healthy cells (proliferating or nonproliferating) have little incentive to maximize the lactate production. Thus, we keep $b$ at 0. Proliferating cells have heavy ATP rate objective while nonproliferating cells have heavy ATP yield objective. Thus, we vary $a$ from 0.5 to 1, and set $c = 1 - a$ to capture the behaviors of proliferating and nonproliferating cells. Other parameter values are from the base case variant in Section 3.3: $\alpha_1 = \alpha_2 = 0.5$, $\alpha_3 = 25$, and $\mathcal{T} = 200$. ATP yield and ATP rate are conflicting objectives with this parameter setting.

Depending on the value of $V_{glucose}$, the feasible region can be either a quadrilateral or a triangle (Fig. 8a and 8b). $f_{fermentation}$ and optimum locations corresponding to different glucose availabilities and objective functions are in Fig. 8c. Data points corresponding to two selected objective weight sets ($a = 0.7$, $b = 0$, $c = 0.3$; $a = 1$, $b = 0$, $c = 0$) (grey lines in Fig. 8c) are plotted in Fig. 8d to show the precise $f_{fermentation}$ values.

A heavy ATP yield objective ($c > 0.42$ or equivalently $a < 0.58$) drives the cells to use the respiration pathway regardless of the glucose availability. The optimal solution for 100% respiration is at point B or H in Fig. 8a and 8b. A heavy ATP production rate objective ($c < 0.42$ or equivalently $a > 0.58$) leads to interesting cellular behaviors dependent on $V_{glucose}$. Severely limited glucose availability ($V_{glucose} < 7.84$) forces cells to use 100% respiration. Moderate glucose availability drives cells to use 100% respiration or mixed respiro-fermentation, depending on the exact values of the objective weights. Mixed respiro-fermentation corresponds to point F in Fig. 8a. Abundant glucose ($V_{glucose} > 200$) enables cells to use 100% fermentation.

## 4 Discussion

Previously, Schuster et al. (2015) and Möller et al. (2018) proposed a minimal model that explains the origin of the Warburg Effect from the energy perspective (i.e. ATP). To accomplish this, they varied cellular parameters (e.g. $\alpha_3$) and kept the same objective function of maximizing the ATP production rate. In this work, we extend the applicability of their model to multiple cell types by using the multi-objective optimization approach. Different cell types also show different extents of the Warburg Effect (Sun et al., 2019; Vander Heiden et al., 2009; Warburg, 1956). Leveraging our NLMOFBA model, we show that such phenomena can be explained by different cellular objectives. Concretely, NLMOFBA connects the cellular pathway preference with the cellular objectives by overlaying the cellular operating modes on the cellular objective map. When investigating the impact of cellular objectives, we fix other cellular parameters whenever possible. Thus, in these case studies, the variations of the cellular behaviors are only due to different cells' objectives captured by the computation model. Our model indicates that healthy nonproliferating cells almost always use respiration because of their objective to maximize the ATP yield, and Sertoli cells almost always use fermentation because of their objective to maximize the lactate production rate (Fig. 4, 5 and 6). Because of the ATP rate objective, cancer cells and healthy proliferating cells could use either respiration or fermentation, depending on the parameter settings and their exact objective weight combinations (Fig. 4, 5 and 6). Thus, we successfully show that the cellular objective can explain the Warburg Effect in different cell types (Liberti and Locasale, 2016; Oliveira et al., 2014; Sun et al., 2019; Vander Heiden et al., 2009; Warburg, 1956).

Besides being capable of modelling multiple cell types, NLMOFBA can also output other biologically significant results consistent with the biology literature. For example, Fig. 7b is consistent with the fact that the compromised mitochondria function could result in the Warburg Effect (Harris & Johnson, 2019). Fig. 8c is supported by the observation that some cells use respiration when glucose is severely limited, but they switch to mixed respiro-fermentation or fermentation when glucose becomes more available (Vander Heiden et al., 2009). In this computation model, increasing glucose availability is mathematically equivalent to an upregulation of glycolytic enzymes. Both changes can be modeled by raising the upper bound on the glycolysis rate. Thus, Fig. 8c can also be supported by another established theory that an upregulation of glycolytic enzymes could lead to the Warburg Effect (Asare-Werehene et al., 2019). Besides being significant in modelling the Warburg Effect, our multi-objective optimization approach can also be leveraged in modelling other cellular phenomena.

Finally, this paper provides a general guideline to incorporate the nonlinear ATP yield term into FBA and avoid the high computational cost associated with nonconvex optimization. This is especially useful when a more complicated metabolic network and more reaction rates are involved.

## 5 Conclusion

We proposed NLMOFBA, a multi-objective

optimization model that explains the impact of cellular objectives on the Warburg Effect in different cell types. In addition, using NLMOFBA, we obtained other important results that are generally consistent with the biology literature. We expect that our model can help readers understand the complicate Warburg Effect. One future direction is to include more reactions (e.g. glutaminolysis) so that the model can produce more biological results and guide the experimental research. Also, it will be interesting to investigate the impact of other cellular objectives such as biomass production on the cellular pathway preference.

## ACKNOWLEDGMENT

No funding was received to assit with the preapartion of this manuscript. The authors have no relevant financial or non-financial interests to disclose.

# APPENDIX

## APPENDIX I
## SUPPORTING FIGURES

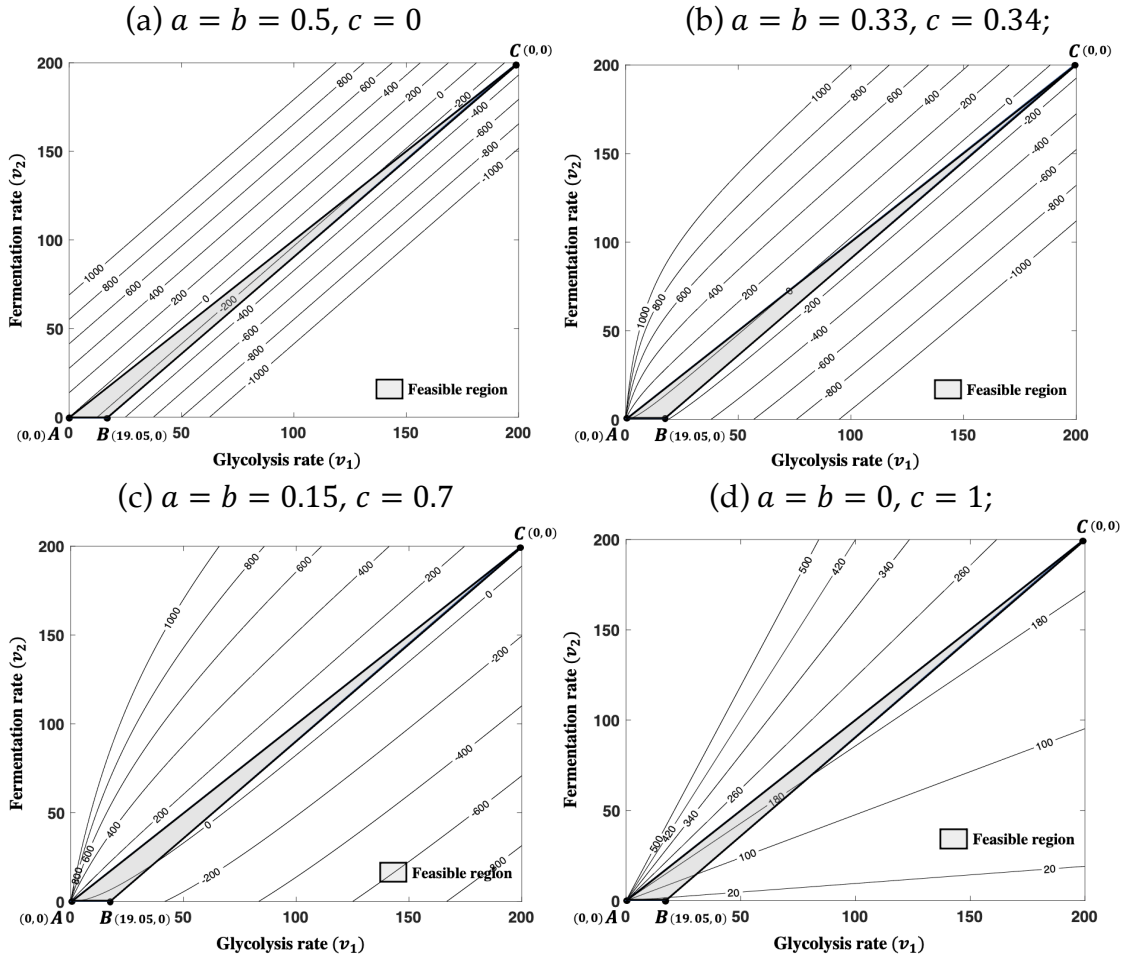

Fig. S1. Level curves of the objective function (in (1)) for four sets of objective weights: (a) $a = b = 0.5, c = 0$; (b) $a = b = 0.33, c = 0.34$; (c) $a = b = 0.15, c = 0.7$; and (d) $a = b = 0, c = 1$. Shaded region is the base case feasible region.

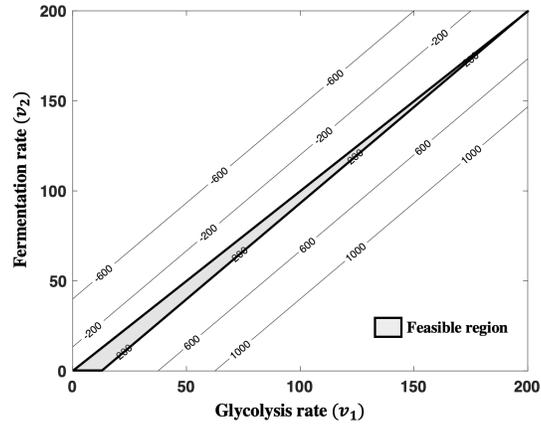

Fig. S2. Level curves of ATP production rate $v_{ATP}$ (in (7)). $\alpha_1 = \alpha_2 = 0.5$, $\alpha_3 = 15.5$ and $\mathcal{T} = 200$. Note that the maximal $v_{ATP}$ (= 200) occurs along the entire feasible region right boundary, which is parallel to the level curves.

# APPENDIX II
# PROOF AND DERIVATION

## 1 Proof of Nonconvexity

**Result.** The objective function

$$F_{objective} = -32av_1 - (b - 30a)v_2 + 210(1 - a - b)\frac{v_2}{v_1}$$

is generally nonconvex over the feasible regions.

**Proof.**

**Theorem 1.** A twice differentiable function is convex if and only if its hessian matrix is positive semidefinite.

The objective function $F_{objective}$ is twice differentiable when $v_1 \in (0, \infty)$.

**Theorem 2.** A symmetric matrix is positive semidefinite if and only if all of its eigenvalues are nonnegative.

First, we find the hessian matrix of the objective function:

$$H = \begin{bmatrix} \frac{d^2 F_{objective}}{(dv_1)^2}, & \frac{d^2 F_{objective}}{dv_1 dv_2} \\ \frac{d^2 F_{objective}}{dv_2 dv_1}, & \frac{d^2 F_{objective}}{(dv_2)^2} \end{bmatrix} = \begin{bmatrix} 420(1 - a - b)v_2 v_1^{-3}, & -210(1 - a - b)v_1^{-2} \\ -210(1 - a - b)v_1^{-2}, & 0 \end{bmatrix}$$

Denote $H$ as:

$$H = \begin{bmatrix} \varphi, & \gamma \\ \gamma, & 0 \end{bmatrix}$$

Two eigenvalues of $H$ are:

$$\lambda_1 = \frac{\varphi + \sqrt{\varphi^2 + 4\gamma^2}}{2}$$

$$\lambda_2 = \frac{\varphi - \sqrt{\varphi^2 + 4\gamma^2}}{2}$$

Since $\varphi$, $\varphi^2$, and $4\gamma^2$ are all non-negative, we know:

$$\lambda_1 \geqslant 0$$

$$\lambda_2 \leqslant 0$$

Also, we know:

$$\lambda_2 = 0 \text{ if and only if } 4\gamma^2 = 4(-210(1-a-b)v_1^{-2})^2 = 0$$

Since $v_1^{-2} \neq 0$,

$$4\gamma^2 = 0 \text{ if and only if } c = 1 - a - b = 0$$

Based on **Theorem 1** and **2**, we conclude that the objective function is nonconvex over the feasible region unless $c = 0$. If $c = 0$, the nonlinear ATP yield term in the objective function vanishes and our model becomes linear programming, which is always convex. □

## 2 Proof of Result 1

**Result 1.** The minimum of the objective function

$$F_{objective} = -32av_1 - (b - 30a)v_2 + 210(1 - a - b)\frac{v_2}{v_1}$$

always occurs at the boundary of the feasible region unless $\frac{dF_{objective}}{dv_2} = 0$ at $(v_{1,opt}, v_{2,opt})$.

**Proof.**

$$\frac{dF_{objective}}{dv_2} = -(b - 30a) + 210(1 - a - b) v_1^{-1}$$

At any fixed $v_1$, $\frac{dF_{objective}}{dv_2}$ is a constant. If $\frac{dF_{objective}}{dv_2} \neq 0$, $F_{objective}$ increases or decreases linearly along the $v_2$ direction. Thus, if $\frac{dF_{objective}}{dv_2} \neq 0$ at the optimal point $(v_{1,opt}, v_{2,opt})$, the optimal point is always on the boundary of the feasible region. If $\frac{dF_{objective}}{dv_2} = 0$ at $(v_{1,opt}, v_{2,opt})$, there could be infinitely many optimal solutions inside the feasible region. □

## 3 Proof of Result 2

**Result 2.** The minimum of the objective function

$$F_{objective} = -32av_1 - (b - 30a)v_2 + 210(1 - a - b)\frac{v_2}{v_1}$$

is always on the right side of the feasible region unless $a = 0$ and $v_2 = 0$.

**Proof.**

$$\frac{dF_{objective}}{dv_1} = -32a - 210(1 - a - b)v_2 v_1^{-2}$$

Quantities $a$, $1 - a - b$, $v_2$, and $v_1^{-2}$ are nonnegative. We know:

$$\frac{dF_{objective}}{dv_1} \leqslant 0$$

$\frac{dF_{objective}}{dv_1} = 0$ if and only if at least one of the two following conditions holds:

1. $a = 0$, $b = 1$, and $c = 0$

2. $a = 0$ and $v_2 = 0$

If none of two conditions holds, $\frac{dF_{objective}}{dv_1} < 0$. $F_{objective}$ decreases as $v_1$ increases. As a result, at each $v_2$, the minimum of $F_{objective}$ always occurs on the right side of the feasible region. If $a = 0$, $b = 1$, and $c = 0$, the cells aim to maximize the lactate production only. The optimal solution corresponding to this objective weight combination always occurs at the intersection of Constraint (2) and Constraint (5) (point C in Fig. 3). Point C is the intersection of the left and right sides of the feasible region, and it is also on the right side. Thus, unless $a = 0$ and $v_2 = 0$, the optimal point is always on the right side of the feasible region. □

## 4 Derivation of the Interface Equation

**Interface equation:**

$$a = \left(\frac{\alpha_1 \mathcal{T} + \alpha_3 \mathcal{T} + 210(\alpha_1+\alpha_3)(\alpha_1+\alpha_2)}{30\alpha_1 \mathcal{T} + 32\alpha_2 \mathcal{T} - 2\alpha_3 \mathcal{T} - 210(\alpha_1+\alpha_3)(\alpha_1+\alpha_2)}\right) b - \frac{210(\alpha_1+\alpha_3)(\alpha_1+\alpha_2)}{30\alpha_1 \mathcal{T} + 32\alpha_2 \mathcal{T} - 2\alpha_3 \mathcal{T} - 210(\alpha_1+\alpha_3)(\alpha_1+\alpha_2)}$$

**Derivation.** The derivation of the interface equation relies on the assumption that cells are either using 100% fermentation or using 100% respiration. To ensure that this assumption is valid in the cases investigated by us, all the interfaces in Fig. 6 are verified by running NLMOFBA.
At the interface, the uses of 100% respiration and 100% fermentation result in the same objective value. Thus, we have:

$$F_{objective, 100\% \; fermentaion} = F_{objective, 100\% \; respiration}$$

**Result 2** indicates that the optimum is at the right boundary of the feasible region if $a \in (0, 1]$. Since the right boundary of the feasible region is the total cellular enzyme resource, we know all cellular resource will be depleted when $a \in (0, 1]$.

If cells use 100% respiration and deplete the cellular enzyme resource, we obtain:

$$\begin{cases} v_1 = v_3 \\ v_2 = 0 \\ \alpha_1 v_1 + \alpha_2 v_2 + \alpha_3 v_3 = \mathcal{T} \end{cases}$$

Solve the system of equations and we get:

$$\begin{cases} v_1 = \mathcal{T}/(\alpha_1+\alpha_3) \\ v_2 = 0 \\ v_3 = \mathcal{T}/(\alpha_1+\alpha_3) \end{cases}$$

Plug the expression of $v_1$, $v_2$ and $v_3$ into the objective function and we get:

$$F_{objective, 100\% \; respiration} = -\frac{32a\mathcal{T}}{\alpha_1+\alpha_3}$$

If cells use 100% fermentation and deplete the cellular enzyme resource, we obtain:

$$\begin{cases} v_1 = v_2 \\ v_3 = 0 \\ \alpha_1 v_1 + \alpha_2 v_2 + \alpha_3 v_3 = \mathcal{T} \end{cases}$$

Solving this system of equations results in:

$$\begin{cases} v_1 = \mathcal{T}/(\alpha_1+\alpha_2) \\ v_2 = \mathcal{T}/(\alpha_1+\alpha_2) \\ v_3 = 0 \end{cases} \tag{A1}$$

Plug the expression of $v_1$, $v_2$ and $v_3$ into the objective function and we get:

$$F_{objective,100\% \; fermentation} = -\frac{32a\mathcal{T}}{\alpha_1+\alpha_2} - \frac{(b-30a)\mathcal{T}}{\alpha_1+\alpha_2} + 210(1-a-b)$$

By equating $F_{objective,100\% \; respiration}$ and $F_{objective,100\% \; fermentation}$, we get:

$$-\frac{32a\mathcal{T}}{\alpha_1+\alpha_3} = -\frac{32a\mathcal{T}}{\alpha_1+\alpha_2} - \frac{(b-30a)\mathcal{T}}{\alpha_1+\alpha_2} + 210(1-a-b)$$

After rearranging this equation, we get the interface equation:

$$a = \left(\frac{\alpha_1\mathcal{T}+\alpha_3\mathcal{T}+210(\alpha_1+\alpha_3)(\alpha_1+\alpha_2)}{30\alpha_1\mathcal{T}+32\alpha_2\mathcal{T}-2\alpha_3\mathcal{T}-210(\alpha_1+\alpha_3)(\alpha_1+\alpha_2)}\right)b - \frac{210(\alpha_1+\alpha_3)(\alpha_1+\alpha_2)}{30\alpha_1\mathcal{T}+32\alpha_2\mathcal{T}-2\alpha_3\mathcal{T}-210(\alpha_1+\alpha_3)(\alpha_1+\alpha_2)}$$

So far, we have shown that the interface equation is valid if $a \in (0, 1]$. If $a = 0$, cells may not deplete all the cellular enzyme resource. If $a = 0$, $F_{objective}$ becomes:

$$F_{objective,a=0} = -bv_2 + 210(1-b)\frac{v_2}{v_1} \tag{A2}$$

If cells use 100% respiration, the variable $v_2$ is 0. Thus, the value of $F_{objective,100\% \; respiration,a=0}$ is 0 regardless of $v_1$. If cells use 100% fermentation, we know $v_2 \neq 0$. **Result 2** tells us that when $a = 0$ and $v_2 \neq 0$, the cellular enzyme resource is depleted. We have derived (A1) when the cellular enzyme resource is depleted, and cells use 100% fermentation. Plug the expressions of $v_1$ and $v_2$ in (A1) into (A2), we get

$$F_{objective,100\% \; fermentation,a=0} = -\frac{b\mathcal{T}}{\alpha_1+\alpha_2} + 210(1-b)$$

After equating $F_{objective,100\% \; respiration,a=0}$ and $F_{objective,100\% \; fermentation,a=0}$, we get

$$-\frac{b\mathcal{T}}{\alpha_1+\alpha_2} + 210(1-b) = 0$$

After rearranging this equation, we get:

$$b = \frac{210(\alpha_1+\alpha_2)}{\mathcal{T}+210(\alpha_1+\alpha_2)}$$

The point $(\frac{210(\alpha_1+\alpha_2)}{\mathcal{T}+210(\alpha_1+\alpha_2)}, 0)$ is on the interface equation. Thus, the interface equation is valid for $a \in [0, 1]$ and $b \in [0, 1]$.